\documentclass[a4paper,twoside]{article}
\usepackage{epsfig}
\usepackage{subcaption}
\usepackage{calc}
\usepackage{amssymb}
\usepackage{amstext}
\usepackage{amsmath}
\usepackage{amsthm}
\usepackage{multicol}
\usepackage{pslatex}
\usepackage{apalike}
\usepackage{algorithm2e,url}
\usepackage[bottom]{footmisc}
\usepackage{SCITEPRESS}     % Please add other packages that you may need BEFORE the SCITEPRESS.sty package.
\usepackage{xcolor}

\begin{document}

\title{On Addressing Isolation in Blockchain-Based Self-Sovereign Identity}

\author{\authorname{Andreea Elena Drăgnoiu
 \sup{1,2}\orcidAuthor{0000-0003-0591-3135}, Andrei Ciobanu\sup{1,2} and Ruxandra F. Olimid\sup{1,2}\orcidAuthor{0000-0003-3563-9851}
}
 \affiliation{\sup{1}Department of Computer Science, University of Bucharest, Bucharest, Romania}
 \affiliation{\sup{2} Research Institute of the University of Bucharest (ICUB), Bucharest, Romania}
 \email{andreea-elena.panait@drd.unibuc.ro, andrei.ciobanu10@s.unibuc.ro, ruxandra.olimid@fmi.unibuc.ro}
 }

%\author{Author}{Blinded for review}{}{}{}

% \author{Andreea Elena Drăgnoiu}{Department of Computer Science, University of Bucharest, Bucharest, Romania \and Research Institute of the University of Bucharest (ICUB), Bucharest, Romania \and \url{https://www.linkedin.com/in/andreea-elena-dragnoiu} }{andreea-elena.panait@drd.unibuc.ro}{https://orcid.org/0000-0003-0591-3135}{This work was supported by a grant of the Ministry of Research, Innovation and Digitalization, CNCS/CCCDI - UEFISCDI, project number ERANET-CHISTERA-IV-PATTERN, within PNCDI IV.}

% \author{Andrei Ciobanu}{Department of Computer Science, University of Bucharest, Bucharest, Romania \and Research Institute of the University of Bucharest (ICUB), Bucharest, Romania \and  }{andrei.ciobanu10@s.unibuc.ro}{}{This work was supported by a grant of the Ministry of Research, Innovation and Digitalization, CNCS/CCCDI - UEFISCDI, project number ERANET-CHISTERA-IV-PATTERN, within PNCDI IV.}

% \author{Ruxandra F. Olimid}{Department of Computer Science, University of Bucharest, Bucharest, Romania \and Research Institute of the University of Bucharest (ICUB), Bucharest, Romania \and \url{https://unibuc.ro/user/ruxandra.olimid/?lang=en} }{ruxandra.olimid@fmi.unibuc.ro}{https://orcid.org/0000-0003-3563-9851}{This work was supported by a grant of the Ministry of Research, Innovation and Digitalization, CNCS/CCCDI - UEFISCDI, project number ERANET-CHISTERA-IV-PATTERN, within PNCDI IV.}

%\authorrunning{A.E.Drăgnoiu, A.Ciobanu and R.F.Olimid} 

\keywords{Self-Sovereign Identity (SSI), Blockchain, Cross-chain, Digital identity}

\abstract{Self-Sovereign Identity (SSI) grants holders full ownership and control of their digital identities, being the ultimate digital identity model. Operating in a decentralized manner, SSI enables the verification of claims, including privacy-preserving mechanisms.
Blockchain, which can be used to implement a Verifiable Data Registry (VDR), is often considered one of the pillars of SSI, along with Decentralized Identifiers (DIDs) and Verifiable Credentials (VCs). Unfortunately, blockchains are mostly siloed, affecting the interoperability and universality of SSI.
We investigate the effect of blockchain isolation on blockchain-based SSI. We first define possible scenarios for cross-chain SSI and exemplify with real-life use cases. We then define specific requirements for cross-chain SSI and identify challenges, also in relation to the identified scenarios. We explore various solutions to achieve blockchain interoperability, with a focus on SSI. In particular, we identify the advantages and disadvantages of distinct cross-chain models for cross-chain SSI. 
Finally, we address the usability of cross-chain SSI and discuss security and privacy aspects, opening the way for future research.}

\onecolumn \maketitle \normalsize \setcounter{footnote}{0} \vfill

\section{Introduction}
\label{sec:intro}

\bigskip
% A digital identity represents an entity (e.g., a person, a device, an organization) in the digital world. Lately, the idea of implementing digital identity on a large scale has become of great importance in Europe (e.g., with the adoption of the EU Digital Identity Wallet \cite{eu_id_wallet} or the EU Age Verification Solution \cite{eu_age_solution}) and the rest of the world (e.g., \cite{medium_astrakode,mazzocca2025survey} list digital identity systems and regulations in force in America, Africa, Asia, Australia and Oceania). Recently, the digital identity has demonstrated its usability and ease for travel \cite{ewc_travel}, a key use case for EU Digital Identity Wallets \cite{ec_use_cases}.

% Conceptually and technologically, digital identities can be approached in several ways. A relatively new model called \textit{Self-Sovereign Identity (SSI)} gives users full control of their identities without the need to rely on external providers, thus reducing risks such as identity theft. SSI is a digital identity model in which users fully own, define, and manage their identities without relying on third parties \cite{dragnoiu2024identitymanagementsolutionarweave}. Moreover, in the new paradigm of SSI, the users can disclose identification data based on a given context, in a selective and privacy-preserving way. This is achievable by three pillars, namely: (1) Distributed IDentifier (DID), (2) Verifiable Data Registry (VDR), and (3) Verifiable Credential (VC) \cite{zecchini2023building}.

A digital identity represents an entity, such as a person, device, or organization, in the digital world. Recently, large-scale digital identity initiatives have gained prominence, especially in Europe with the EU Digital Identity Wallet \cite{eu_id_wallet} and the EU Age Verification Solution \cite{eu_age_solution}, and globally in various regions \cite{medium_astrakode,mazzocca2025survey}. Digital identities have also proven to be useful for travel \cite{ewc_travel}, a key use case for the EU Digital Identity Wallet \cite{ec_use_cases}.

Among the different digital identity models, Self-Sovereign Identity (SSI) stands out as giving users full control over their identities without relying on third parties, thereby reducing risks such as identity theft \cite{dragnoiu2024identitymanagementsolutionarweave}. SSI allows users to share only the necessary data in a privacy-preserving way, based on context. It relies on three main components \cite{zecchini2023building}: (1) Distributed Identifier (DID), (2) Verifiable Data Registry (VDR), and (3) Verifiable Credential (VC).

%-------------------------------------------
%-------------------------------------------
%-------------------------------------------
%\subsection{Motivation}
%\label{sec:intro_motivation}

% One specific implementation of VDR, considered to bring the most decentralization of DIDs, is a blockchain \cite{dif_faq}. By using a blockchain, SSI eliminates the need for central authorities and avoids users exposing their sensitive information to a single entity \cite{mazzocca2025survey}.
% Blockchains come in different flavors, each targeting a specific functionality or goal (which in turn, introduces specificity and makes them operate in silos) \cite{sidhu2022trust,zecchini2023building}. To get the advantage of most of those, for a particular use case, decentralized applications (dApps) that run over more than a single blockchain have emerged as a natural solution. In conclusion, users should interact with multiple blockchains and other users, possibly placing their identities in different blockchains~\cite{zecchini2023building}.

\textbf{Motivation.}
Blockchain is a key VDR implementation that provides a high level of DID decentralization and eliminates the need for central authorities \cite{dif_faq,mazzocca2025survey}. Different blockchains serve specific purposes and often operate in silos \cite{sidhu2022trust,zecchini2023building}, so decentralized applications (dApps) spanning multiple chains have emerged, allowing users to manage identities across blockchains \cite{zecchini2023building}.
%In particular, one user can create multiple identities, on the same or sometimes distinct blockchains, to achieve his/her purposes \cite{zecchini2023building}. 

% In this context, the necessity of cross-chain communication comes in different flavors, including: (a) a user that creates an identity on one blockchain and needs to use functionalities of another blockchain, (b) exchange or merging of identities that belong to a single user but are anchored in distinct blockchains, and (c) interactions among distinct users defined in different blockchains.
% Real use-cases that exemplify such scenarios include the cross-chain verification of certificates and diplomas \cite{tran2024crosscert,tan2023verification,boulet2025digital}, the cross-border healthcare data \cite{broshka2024evaluating}, making travel easier by simplifying the check-in and controls procedures \cite{misc_protokol,ec_use_cases}, or interoperability across different
% blockchain-based supply chain systems by verifying asset alterations tracked across multiple blockchains \cite{mezquita2023blockchain}. In this context, it is notable to mention EBSI, the European Blockchain Services Infrastructure (EBSI) \cite{ebsi}, which aims to leverage blockchain for cross-border services, and the Decentralized Identity Foundation (DIF) \cite{dif}, which is looking towards established interoperable and global standards. For example, the integration of EBSI with similar ledgers belonging to other countries would extend interoperability worldwide.

In this context, cross-chain communication is needed for: (a) using an identity from one blockchain on another, (b) exchange or merging of identities that belong to a single user but are anchored in distinct blockchains, and (c) interactions among users on different blockchains. Real use cases include cross-chain certificate verification \cite{tran2024crosscert,tan2023verification,boulet2025digital}, cross-border healthcare data \cite{broshka2024evaluating}, travel check-ins \cite{misc_protokol,ec_use_cases}, and supply chain interoperability \cite{mezquita2023blockchain}. Initiatives such as the Decentralized Identity Foundation (DIF) \cite{dif} and the European Blockchain Services Infrastructure (EBSI) \cite{ebsi} aim to standardize and extend cross-chain interoperability globally.

SSI should be highly interoperable with most service providers \cite{yildiz2023toward}.
Although many advances in standardization have been made recently, e.g., by W3C \cite{w3c_did,w3c_did_draft,w3c_verifcred}, there are still many aspects to consider. In particular, there is a lack of standardization on how holders can link their identities defined in distinct blockchains and transfer related data between ledgers \cite{zecchini2023building}. 
According to \cite{zhong2021jointcloud}, DIDs can only be verified on a single blockchain; therefore, their interoperability across different blockchains is an explicit limitation.
%\textit{"DIDs can only conduct verification within a single blockchain, which limits the interoperability of DIDs on different blockchains"}.
With blockchain considered to bring the highest decentralization of DIDs, and with DIDs adoption at a larger scale, these aspects require further investigation.

%\subsection{Contributions}
%\label{sec:intro_contribution}

\textbf{Contributions.}
Our contribution is summarized as follows. (a) We define possible scenarios for cross-chain blockchain-based SSI and exemplify with use-cases. (b) We list requirements for cross-chain SSI, with a focus on security and privacy, but also interoperability and consistency. (c) We identify challenges for cross-chain SSI, also in relation to the specifics of the identified scenarios. (d) We discuss various models for realizing cross-chain, identifying their weaknesses and strengths in the context of SSI, and we refer to existing projects and industry solutions. (e)~We address the usability of cross-chain SSI and discuss security and privacy aspects; we formulate open questions and thus open the way for future work.

%Our contributions are summarized as follows: (a) defining cross-chain SSI scenarios with real use cases, (b) outlining requirements focused on security, privacy, interoperability, and consistency, (c) identifying challenges related to these scenarios, (d) analyzing cross-chain models with their strengths and weaknesses and referencing existing projects, and (e) discussing usability, security, and privacy while highlighting open questions for future research.

%-------------------------------------------
%-------------------------------------------
%-------------------------------------------
%\subsection{Outline}
%\label{sec:intro_outline}

% The paper is organized as follows. Section \ref{sec:back} gives the preliminaries, including SSI and cross-chain communication. Section \ref{sec:related_works} presents the existing work. Section \ref{sec:cc_ssi} introduces the identified scenarios for SSI in the context of cross-chain communication, exemplifies with specific use-cases, and defines specific requirements for cross-chain SSI.
% Section \ref{sec:challanges} presents the challenges introduced by blockchain isolation in general and in relation to the identified scenarios. 
% Section \ref{sec:sol_types} examines various cross-chain solutions and discusses their role in cross-chain SSI. Section \ref{sec:disc} discusses different aspects related to cross-chain SSI, with a focus on usability, security, and privacy. The last section concludes.

\textbf{Outline.}
The paper is structured as follows. Section \ref{sec:back} introduces SSI and cross-chain communication. Section \ref{sec:related_works} reviews the related work. Section \ref{sec:cc_ssi} presents the identified scenarios for SSI in the context of cross-chain communication, exemplified by specific use cases, and defines the corresponding requirements. Section~\ref{sec:challanges} discusses the challenges caused by blockchain isolation. Section \ref{sec:sol_types} examines various cross-chain solutions and discusses their role in cross-chain SSI. Section \ref{sec:disc} discusses different aspects of cross-chain SSI, with a focus on usability, security, and privacy. The last section concludes.

%-------------------------------------------
%-------------------------------------------
%-------------------------------------------
\section{Preliminaries}
\label{sec:back}

%-------------------------------------------
%-------------------------------------------
%-------------------------------------------
%\subsection{Decentralized Digital Identity}
%\label{sec:back_digital_id}

%\ruxandra{\textbf{@andreea:} Please go over this section again so that we agree on a final version}

%-------------------------------------------
%-------------------------------------------
%-------------------------------------------
\subsection{Decentralized IDentifier (DID)}
\label{sec:back_did}

A \textit{Decentralized Identifier (DID)} is a globally unique identifier that enables verifiable and decentralized digital identity management. A DID represents an entity (e.g., an individual, an organization, a digital object), which is referred to as the \textit{DID subject}. 
A DID has a universal Uniform Resource Identifier (URI) structure formed of three parts: (1) a required \textit{did:} prefix, (2) a \textit{method:} an identifier that specifies the DID method, and (3) a unique, method-specific identifier for the subject. Each blockchain introduces its own method for constructing a DID. Table \ref{tab:did_method} exemplifies different blockchains and their specific DID methods. A more complete list of DID methods is available at \cite{w3c_did_methods}.

%---------------------------------
\begin{table*}
\begin{footnotesize}
\begin{center}
\begin{tabular}{p{0.25\textwidth}p{0.15\textwidth}p{0.4\textwidth}}
\hline
\textbf{Blockchain} & 
\textbf{DID Method} &
\textbf{Example DID}\\
\hline
Ethereum & did:ethr & did:ethr:0x123456789abcde... \\
Bitcoin & did:ion & did:ion:EiD72hVo1K8...\\
%Solana & did:sol & did:sol:5F8qzPn7eVhHyhVvM... \\
%Polkadot / Substrate & did:kilt & did:kilt:4rGrefhSYBq5Pfh...\\
Polygon & did:polygonid & did:polygonid:polygon:mumbai:2q3d9fsfs...\\
%Tezos & did:tz & did:tz:tz1VSUr8wwNhLAzempoc...\\
Other (General) & did:$<$key$>$ & did:$<$key$>$:z6Mkwdasdgswijsdj...\\
\hline
\end{tabular}
\end{center}
\end{footnotesize}
\caption{Examples of DIDs over different blockchains}
\label{tab:did_method}
\end{table*}
%---------------------------------

\begin{figure}[b!]
\centering
\includegraphics[width=0.45\textwidth]{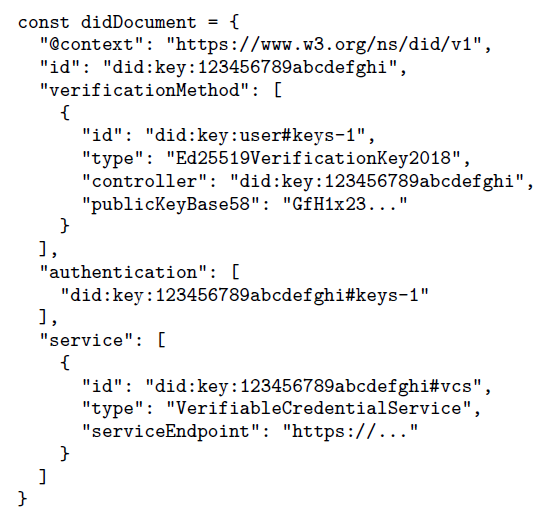}
\caption{Example of DID document~\cite{dragnoiu2024identitymanagementsolutionarweave,w3c_did}}
\label{fig_did_document}
\centering
\end{figure}

A DID uniquely identifies a \textit{DID document} that, besides the DID and the DID subject, contains associated information such as public keys, service endpoints, and authentication protocols. A \textit{DID controller} (usually the DID subject) is authorized to update this document, as specified by the \textit{DID method} 
%The controller may be the DID subject, but can also be a separate entity. This distinction is essential for handling key loss or compromise. It is possible to define multiple controllers for a single DID document
~\cite{w3c_did,w3c_did_draft}. Figure \ref{fig_did_document} gives an example of a DID document in JSON format.
The W3C standardizes DIDs through the DID Core, which defines the architecture, data model, and representations~\cite{w3c_did}. At the time of writing, a new version is defined as a draft~ \cite{w3c_did_draft}. %This standard is implementation-agnostic and can be applied to decentralized storage networks, where DID documents can be created, stored, and resolved using standard DID libraries.

%---------------------------------
%---------------------------

%-------------------------------------------
%-------------------------------------------
%-------------------------------------------
\subsection{Verifiable Credential (VC)}
\label{sec:back_VC}

%---------------------------------
\begin{figure*}
\centering
\includegraphics[width=0.75\textwidth]{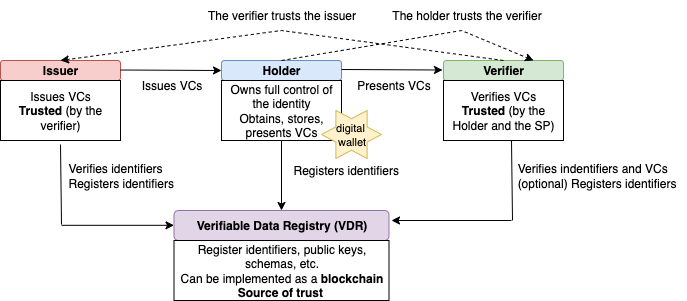}
\caption{The VC ecosystem and SSI trust framework overview (adapted from \cite{w3c_verifcred,ebsi_vc})}
\label{fig_vc_framework}
\centering
\end{figure*}

Verifiable Credentials (VCs) are digital, privacy-preserving proofs of claims about an entity. A VC contains a subject, claims, (cryptographic) proofs, and metadata such as as the issuer, issuance date, and expiration date \cite{w3c_verifcred}. Claims might state, for example, a subject’s age or nationality, with the proof to ensure authenticity.

Key entities include the \textit{issuer} (who creates and signs the VC), the \textit{holder} (who owns the VC and can present it), and the \textit{verifier} (who validates the presentation). %Presentations can merge VCs and selectively reveal claims, but here we focus on VCs themselves. 
Verification is usually mediated via a Verifiable Data Registry (VDR), which can be a blockchain, distributed ledger, or trusted database. While DIDs can also use non-blockchain systems (e.g., Key Event Receipt Infrastructure (KERI), InterPlanetary File System (IPFS), and Public Key Infrastructure (PKI) \cite{yildiz2023toward}), this work focuses solely on blockchain-anchored DIDs. Figure \ref{fig_vc_framework} illustrates this ecosystem as formalized by the W3C \cite{w3c_verifcred}.
%-------------------------------------------
%-------------------------------------------
%-------------------------------------------
\subsection{Self-Sovereign Identity (SSI)}
\label{sec:back_ssi}

Identity management (IdM) traditionally follows a three-party model of holders, issuers, and verifiers, as defined by the VC framework: holders request access, issuers provide credentials, and verifiers validate them \cite{panait2020identitymanagementblockchain}. Self-Sovereign Identity~(SSI) modernizes this model by giving users full control over their digital identities—allowing them to create, manage, and revoke credentials without intermediaries, reducing risks such as identity theft. SSI relies on DIDs \cite{w3c_did} and VCs \cite{w3c_verifcred}.

In blockchain-based VDRs, DID documents are stored on-chain for integrity, while VCs remain off-chain for privacy \cite{dragnoiu2024identitymanagementsolutionarweave,yildiz2023toward,zecchini2023building}, requiring trust between issuers and verifiers. Two SSI models are typically used \cite{panait2020identitymanagementblockchain}:
(1) the \textit{Identifier Registry Model}, which stores only identifiers—preserving privacy but offering weaker tamper protection, and
(2) the \textit{Claim Registry Model}, which stores identifiers and claim hashes—enhancing verification but reducing privacy.

SSI underpins Web3 identity, combining VCs, digital wallets for secure credential storage, and blockchains as decentralized trust layers \cite{ebsi_vc}, while aligning with frameworks such as eIDAS \cite{eidas_doc} and GDPR \cite{gdpr}.
%-------------------------------------------
%-------------------------------------------
%-------------------------------------------
\subsection{Cross-Chain Communication}
\label{sec:back_cross_chain}

\textit{Cross-chain} solutions address the isolation of blockchains, enabling interoperability and communication between distinct chains. \textit{Cross-chain communication} enables stand-alone blockchains to exchange data, assets, or state proofs in a secure, verifiable, and ideally efficient manner. Instead of treating each blockchain as a silo, cross-chain communication breaks down isolation, allowing users (and developers) to leverage the unique features of different chains within a single ecosystem.  In the context of SSI, cross-chain solutions aim to make digital identities portable and verifiable across chains.

Cross-chain solutions address blockchain isolation by enabling secure, verifiable interoperability between distinct chains. They allow blockchains to exchange data, assets, or state proofs, making SSI identities portable and verifiable across networks. Various cross-chain technologies exist, including notary schemes, side chains, relay chains, hash locking, blockchain routers, cross-chain smart contracts, and distributed private key technology \cite{xie2022cross,tran2024crosscert,zhong2021jointcloud}, with many studies analyzing their pros and cons and interoperability challenges \cite{ou2022overview,augusto2024sok,zhang2024security,belchior2024brief}. However, a comprehensive taxonomy remains unsettled. We discuss cross-chain SSI solutions in Section \ref{sec:sol_types}.

%-------------------------------------------
%-------------------------------------------
%-------------------------------------------
\section{Related Work}
\label{sec:related_works}

The work on SSI and cross-chain as independent topics is vast and out of our present scope. 
We only discuss joint work and related concepts.

\textbf{Interoperability and portability}. In our context, we can independently argue about interoperability and portability in the context of either SSI or blockchain.
On the SSI side, the paper \cite{yildiz2023toward} discusses SSI interoperability, dissects it into several layers, proposes a fine-grained definition, and considers a reference model that allows addressing interoperability among different implementations. It also serves as a tutorial and knowledge base for SSI interoperability, providing a comprehensive overview of various solutions.  
Other works, such as \cite{gruner2021analyzing}, examined interoperability and portability concepts for SSI before. On the blockchain side, the amount of work on interoperability and portability is vast, with the papers \cite{ou2022overview,augusto2024sok,zhang2024security,belchior2024brief} representing just a few examples. A recent survey on SSI on VC also touches the case of cross-chain interoperability \cite{mazzocca2025survey}.

\textbf{Cross-chain SSI.}
\cite{zecchini2023building} introduces a framework to build a cross-chain identity in the SSI paradigm, from its definition to the data synchronization of multiple blockchain identities. For this, the authors define a new relay called \textit{StateRelay}, aiming at reducing the final usage costs, and compare it with the state-of-the-art. The solution maintains users' autonomy, and it is claimed to be compliant with the W3C DID standardization and the first solution without the use of intermediaries 
%(e.g., additional blockchains or oracles) 
for permissionless blockchains. Previously, SSI cross-chain without intermediaries was considered for permissioned blockchains \cite{ghosh2021decentralized}.
Formerly, other papers on blockchain-based SSI interoperability have been published. 
\cite{zhong2021jointcloud} proposes a solution to conduct cross-chain verification of DIDs by using smart contracts, but does not consider the W3C specifications. %and the solution seems to store sensitive data on-chain. 
\cite{xie2022cross} shortly surveys cross-chain-based distributed digital identity, presenting underlying concepts such as the architecture of distributed digital identity, the VC data model, and comparing different cross-chain technologies (notary mechanism, side chain and relay chain, hash locking, and distributed private key). However, %the discussion on identity cross-chain is not substantial at all;
the paper mainly considers the relay chain-based identity cross-chain technology, briefly discussing existing implementations such as BitXHub.  
 \cite{sidhu2022trust} investigates a blockchain interoperability solution based on a publish-subscribe architecture. In particular, the paper looks into decentralized identifiers and VCs and the integration with Hyperledger Indy \cite{hyperledgerindy}, Aries \cite{hyperledgeraries}, and Ursa \cite{hyperledgerursa} stack projects.
\cite{liu2022research} combines digital identity and cross-chain technology from a different perspective: the authors propose an improved side chain model by enhancing security in the sense that communicating blockchains register to a distributed digital identity claim. 

\textbf{Use-cases.} %Although not necessarily cross-chain (but single chain), the reference \cite{boulet2025digital} offers a nice overview of European projects and solutions for the digitization of certificates and diplomas. To exemplify with a use case in EBSI Vector, 
Cross-chain SSI finds its utility in multiple areas, including the following.
%\cite{tan2023verification} describes the (Belgium-Italy) cross-border digitization of diplomas and points out interoperability issues regarding digital identity systems.
\cite{tran2024crosscert} introduces \textit{CrossCert}, a cross-chain system that enables credential verification between educational and employer blockchains.
%, facilitating anonymity by using zk-proofs. 
\cite{naghmouchi2021automatized} exemplifies the use case of identification and management for smart vehicles, by proposing a cross-chain communication between Ethereum \cite{ethereum} and Hyperledger Indy \cite{hyperledgerindy}. \cite{mezquita2023blockchain} proposes a model for tracking assets in different blockchain-based supply-chain systems in the pharmaceutical sector.

%-------------------------------------------
%-------------------------------------------
%-------------------------------------------
\section{Addressing Cross-Chain SSI}
\label{sec:cc_ssi}

%-------------------------------------------
%-------------------------------------------
%-------------------------------------------
\subsection{Identified Scenarios}
\label{sec:scenarios}

We further define possible scenarios for cross-chain SSI, considering the need for identity and claims recognition across various platforms, including the linking of DIDs and the verification of VCs across different blockchains. Note that by cross-chain SSI, we understand any form of two or more blockchain interoperability in the context of SSI (contrary to the specific understanding of cross-chain SSI in \cite{zecchini2023building}).
Depending on the existence of a DID on each blockchain (Criterion A), we identify two possible scenarios: one single DID is defined on a single blockchain (Scenario A1) or one DID is defined on each blockchain (Scenario A2). 
%We exemplify the two scenarios with a few use-cases.
%Naturally, the cross-chain SSI should also accommodate recognition and verification of VCs across different blockchains. 
Considering the placement of the verifier (Criterion B), the verification could be done on-chain (Scenario~B1) or off-chain (Scenario~B2). By the placement of the verifier on-chain, we understand that the verifier is either running directly on-chain (as a smart contract) or is registered to (at least) one chain. By the placement of the verifier off-chain, we refer to the case in which the verifier is only allowed to read (public) information from the blockchain. We further assume that all blockchains are public, so an off-chain verifier does not need any registration on the blockchain. This might not always illustrate a cross-chain scenario (e.g., the blockchain is just a trusted repository for DIDs, everything else is performed off-chain), so we are interested in the case where the result of the verification will be used on-chain, as a form of a token, attestation, or similar.

\smallskip
\textbf{Criterion A.} \textit{The existence of linked DIDs across distinct blockchains.}

\smallskip
\textbf{- Scenario A1:} The holder owns a single $\mathsf{DID_1}$, which is anchored to a blockchain $\mathsf{BC_1}$, and uses it directly in another blockchain $\mathsf{BC_2}$.

%\textbf{Scenario 1.} \textit{A DID $\mathsf{DID_i}$ that is anchored in a blockchain $\mathsf{BC_i}$ is used directly in another blockchain $\mathsf{BC_j}$, $j \neq i$}.

%In this scenario, $\mathsf{DID_i}$ has no direct correspondent, and it is not linked in any way to a DID anchored in $\mathsf{BC_j}$.
In this scenario, the holder does not necessarily need to own a correspondent $\mathsf{DID_2}$ anchored in $\mathsf{BC_2}$. The generalization to $n \geq 2$ blockchains is straightforward, the holder directly uses the same $\mathsf{DID_1}$ in any blockchain $\mathsf{BC_2}, \dots, \mathsf{BC_n}$.

\smallskip
\textbf{- Scenario A2:} The holder owns a $\mathsf{DID_1}$, which is anchored in a blockchain $\mathsf{BC_1}$, and another $\mathsf{DID_2}$, which is anchored in another blockchain $\mathsf{BC_2}$.
%\textbf{Scenario 2.} \textit{A DID $\mathsf{DID_i}$ that is anchored in a blockchain $\mathsf{BC_i}$ is in correspondence with another DID $\mathsf{DID_j}$ in a distinct blockchain $\mathsf{BC_j}$, $j \neq i$.}

%In this scenario, $\mathsf{DID_i}$ is in direct correspondence with another DID $\mathsf{DID_j}$ that is anchored in the blockchain $\mathsf{BC_j}$.
In this scenario, $\mathsf{DID_1}$ and $\mathsf{DID_2}$ are linked in the sense that they belong to the same holder.
The generalization of this scenario is straightforward and has already been considered in the literature: the holder owns $\{\mathsf{DID_1}, \dots, \mathsf{DID_n}\}$ (the so-called \textit{cross-chain identity} \cite{zecchini2023building}), one DID for each blockchain in $\{\mathsf{BC_1}, \dots \mathsf{BC_n}\}$, $n \geq 2$. 
%Examples of cross-chain identity that illustrate this scenario include \cite{zecchini2023building}.

\smallskip
\textbf{Criterion B.} \textit{The placement of the verifier with respect to the blockchain.}

\medskip
\textbf{- Scenario B1:} The verification of an identity or an identity claim takes place on-chain, on a blockchain $\mathsf{BC_v}$.

In this scenario, the verifier is placed on-chain. Typically, the verification is done through a smart contract. Examples for on-chain identity verification in the SSI context are available at \cite{dragnoiu2024identitymanagementsolutionarweave,zecchini2023building}. 
%Use-cases 1.1 and 1.2 are examples of on-chain verification.

\smallskip
\textbf{- Scenario B2:} The verification of an identity or an identity claim takes place off-chain.

In this scenario, the verifier is placed off-chain, with the blockchain being only used as a tamper-resistant, secure repository for the DID documents. This scenario is less prevalent in blockchain-based IdM, as the digital identity is typically used on-chain rather than off-chain. Nevertheless, it represents a real possibility, for which the blockchain only plays the role of the VDR in the SSI ecosystem. 
%Use-case 2.1 can accommodate both on-chain and off-chain verification (when the facilities residing from the merged assets, rights, reputation, etc. are provided out of a chain, e.g., via a centralized application that requires such corroborative proofs).

\begin{figure*}
\centering
\includegraphics[width=0.5\textwidth]{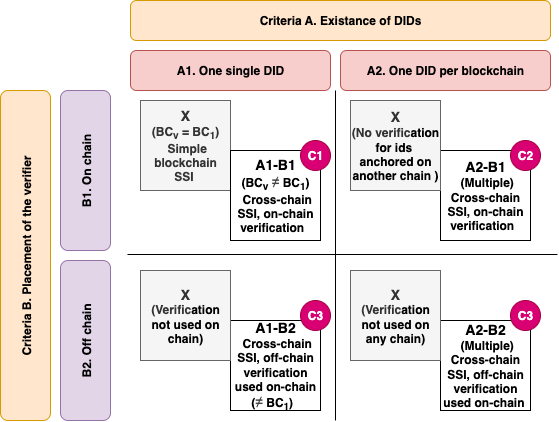}
\caption{Identified scenarios}
\label{fig_scenarios}
\centering
\end{figure*}

\smallskip
% Figure \ref{fig_scenarios} summarizes the identified final scenarios, obtained as all possible combinations of the possibilities mentioned above. To exemplify, A1-B1 is the case where the verifier is placed on a blockchain distinct from the blockchain where the holder's DID is anchored. 
% We are only interested in cross-chain SSI, so we neglect scenarios where the DID is anchored on the same blockchain where the verifier is placed or where the verification is fully off-chain (i.e., the verification result does not affect in any way another blockchain except the one where the DID is anchored). 
% For completeness, we mark these scenarios on the figure too, but mark them with X and further neglect them, as they are out of scope. Lastly, we illustrate by C1-C3 challenges, which we will refer to in Section \ref{sec:challanges}.
Figure \ref{fig_scenarios} summarizes all possible cross-chain SSI scenarios from the combinations above. For example, A1-B1 represents a verifier on a different blockchain than where the holder’s DID is anchored. We exclude cases where the DID and verifier share the same blockchain or where verification is entirely off-chain and mark them with X. Figure \ref{fig_scenarios} also highlights key challenges (C1–C3) discussed in Section \ref{sec:challanges}.

%---------------------------------
%---------------------------

%-------------------------------------------
%-------------------------------------------
%-------------------------------------------
\subsection{Use-Cases}
\label{sec:use-cases}

We further exemplify the previously identified scenarios with use-cases.

\smallskip
\textbf{Use-case 1: Cross-chain authentication/login.}
Using a single SSI identity, a user can access services on multiple blockchains. For this, a cross-chain SSI resolver can verify the user's VC from the anchored blockchain $\mathsf{BC_1}$ and further authenticate the user into the second blockchain $\mathsf{BC_2}$, without the need to duplicate identities. The SSI resolver can take multiple forms, e.g., a bridged DID resolver, a multi-chain DID registry, an off-chain resolver service \cite{li2025blockchain}.
%(refer to Section \ruxandra{TODO} for more details). 
This use case illustrates on-chain verification (Scenario B.1) with no linked DIDs across distinct blockchains (Scenario A.1)

\smallskip
%\ruxandra{@andrei, pls check the text for use-case 2. is it ok? if yes, pls delete the comments and your initial text}
\textbf{Use-case 2: Airdrop token distribution.}
Using a single SSI identity, a user can claim an airdrop token on multiple blockchains. Suppose that the user already has a DID anchored in a blockchain $\mathsf{BC_1}$. When the user wants to claim an airdrop token on $\mathsf{BC_2}$, the user presents its credentials (anchored in $\mathsf{BC_1}$) along with a proof that the token has not been revoked. The verification can be performed by a smart contract on $\mathsf{BC_2}$ that checks the validity of the claim using the user's identity anchored in the initial blockchain $\mathsf{BC_1}$. This way, users can carry a single, reusable identity across chains, while projects on other chains can securely airdrop tokens to verified users without the necessity of storing personal data or issuing credentials for every network. This use case illustrates on-chain verification (Scenario B.1) with no linked DIDs across distinct blockchains (Scenario A.1)

\smallskip
\textbf{Use-case 3: Identity transfer/merging.}
An entity that owns several identities anchored on two (or more) distinct blockchains might want to take advantage of joined assets, rights, or reputation. This can be beneficial in several contexts, including a company merge (initial entities owned identities anchored on distinct blockchains) or considering a privacy-preserving strategy to maintain hidden the presence of a single entity on distinct blockchains. This use case clearly illustrates the necessity of linked DIDs (Scenario A.2) and can accommodate both on-chain (Scenario B.1) and off-chain (Scenario B.2) verification (when the facilities residing from the merged assets, rights, reputation, etc., are provided off-chain, e.g., via a centralized application that requires such corroborative proofs).

\smallskip
Again, we intentionally keep out of scope the scenarios that do not involve cross-chain connectivity. Nevertheless, these scenarios are an independent subject of study. To exemplify with a use-case in EBSI Vector that fits the A1-B1 scenario but only involves a single ledger, \cite{tan2023verification} describes a cross-border use case, pointing out interoperability issues concerning digital identity systems. 

%-------------------------------------------
%-------------------------------------------
%-------------------------------------------
\subsection{Requirements}
\label{sec:requirements}

While the requirements for SSI are well settled \cite{eidas_doc,gdpr,w3c_did,w3c_did_draft,w3c_verifcred,eu_id_wallet,panait2020identitymanagementblockchain,dragnoiu2024identitymanagementsolutionarweave}, for cross-chain SSI there is still a lack in defining the demands. We aim to bridge this gap by outlining the key properties that a cross-chain SSI system should fulfill. Table \ref{tab:requirements} lists these requirements, with a focus on security and privacy aspects. 
%We reiterate that by cross-chain SSI, we understand any form of blockchain interoperability in the context of SSI.

%---------------------------------
\begin{table}
\begin{footnotesize}
%\begin{center}
\begin{tabular}{p{0.22\columnwidth}p{0.65\columnwidth}}
\hline
\textbf{Property} & \textbf{Description} \\
\hline
\textit{Self-sovereignty} & A holder creates and manages its cross-chain identity without (identity) third parties \cite{zecchini2023building} \\
\textit{User consent and control} & Data should not be accessible to other chains without the acceptance and under the control of the holder \\
\textit{Unlinkability} & Two identities $\mathsf{DID_i}$ and $\mathsf{DID_j}$ defined on two distinct blockchains $\mathsf{BC_i}$ and $\mathsf{BC_j}$, respectively, must not be linked to belong to the same holder unless the user declares control of them both \cite{zecchini2023building} \\
\textit{Identity binding} & The holder must provide proof that is in control of each identity $\mathsf{DID}_i$ on blockchain $\mathsf{BC_i}$, $i = 1 \dots n$ \cite{zecchini2023building} \\
\textit{Impersonation resilience} & No (malicious) party that cannot provide proof for $\mathsf{DID_i}$ could act on the holders behalf (in particular, even if one $\mathsf{DID_j}$ is compromised, $\mathsf{DID_i}$, $j \neq i$ is not, as long as the cryptographic proof on $\mathsf{BC_i}$ remains hard)~\cite{zecchini2023building} \\
\textit{Data portability} & The holder can associate or transfer any data linked to the identity $\mathsf{DID_i}$ on $\mathsf{BC_i}$ to an identity $\mathsf{DID_j}$ on $\mathsf{BC_j}$ \cite{zecchini2023building}\\
\textit{Data consistency} & DID-related data (and others) should be consistent on all blockchains (in particular, there should be no conflicts regarding keys renewal, revocation, issuer rights on emitting VCs) \\
\textit{Interoperability} & The solution should be compatible with different (ideally all) blockchains \\ 
\textit{Cross-chain security and privacy} & The cross-chain mechanisms in place should not diminish significantly (ideally, at all) the security and privacy of the overall solution (e.g., should not damage integrity or confidentiality) \\
\textit{Anonymisation} &  The cross-chain mechanisms in place should maintain the anonymization (e.g., identities, interactions, issuing authorities should remain anonymous) \\
\textit{Centralization} & The cross-chain mechanisms in place should not diminish decentralization (in particular, should not introduce centralized intermediaries)\\ 
\textit{Improved user experience} & The implementation of any cross-chain functionality should be as smooth as possible for the end user (e.g., use a single universal wallet) \\
\hline
\end{tabular}
\centering
\end{footnotesize}
\caption{Identified requirements for cross-chain SSI}
\label{tab:requirements}
\end{table}
%---------------------------------

%-------------------------------------------
%-------------------------------------------
%-------------------------------------------
\section{Challenges}
\label{sec:challanges}

%-------------------------------------------
%-------------------------------------------
%-------------------------------------------
\subsection{General Challenges}
\label{sec:challanges_general}

We further discuss the fundamental challenges that arise from the isolated nature of blockchain in relation to blockchain-based SSI.

\smallskip
% Self-sovereignty, by itself, is challenging. In light of blockchains and their isolation, allowing the holder to maintain complete control of his/her identity becomes even more difficult. For example, the revocation of DIDs and DID documents in blockchain-based identity management seems already counterintuitive: the blockchain is built to permanently store information, while the privacy regulations (e.g., GDPR) ask for the user's right to be forgotten. How to manage identity updates and revocations across chains efficiently and scalably remains an open problem. 

% In the SSI model, the holder must create and manage the identity without the need for third parties. In particular, it should work without centralization and intermediaries. However, these are standard solutions for cross-chain communication, and thus, finding alternative acceptable solutions for cross-chain SSI turns out to be not a trivial task. In fact, the first work to claim to achieve this in permissionless blockchains is relatively recent \cite{zecchini2023building}.

% The variety of DID methods \cite{w3c_did_methods}, most of them specific to particular blockchains, increases the difficulty of interoperability. Although they must comply with normative requirements, the W3C does not endorse the DID method or the underlying technology, which allows developers to choose or build their own. This leads to more than 200 complaint DID methods~\cite{w3c_did_methods}.

\textbf{Preserving user-centrism and self-sovereignty.} Preserving user-centrism and self-sovereignty in SSI is challenging, especially across isolated blockchains. Maintaining full control over identities (including updates and revocations) conflicts with blockchain immutability and privacy regulations like GDPR, making cross-chain management an open problem.
SSI requires holders to manage identities without intermediaries, yet typical cross-chain solutions rely on centralization. Achieving decentralized cross-chain SSI is recent and limited \cite{zecchini2023building}. Interoperability is also complex because of the variety of DID methods, mostly blockchain-specific, with over 200 W3C-compliant options \cite{w3c_did_methods}.

\smallskip
\textbf{Security of VCs.}
The security of VCs along their whole life-cycle (creation/issuance, storage, presentation, verification, revocation, etc.) must be considered. In particular, solutions that expose VCs to smart contracts - or, in any way, store sensitive information on-chain (even in protected form) - are condemned to privacy leaks in the future. This holds because, by construction, blockchains' storage is permanent and the data protection mechanisms are ephemeral (the cryptography is mostly computational and vulnerable to quantum attacks)~\cite{panait2020identitymanagementblockchain,dragnoiu2024identitymanagementsolutionarweave}.

\smallskip
\textbf{Standardization and technological limitations.}
Even though many advances have been made recently in the standardization of DID and VC \cite{w3c_did,w3c_did_draft,w3c_verifcred}, the implementation of the VDR is out of their scope. DID methods are often associated with a specific VDR~\cite{w3c_did}, and distinct DID methods can be incompatible. 
Moreover, cross-chain connectivity also lacks standardization, making things even more challenging for SSI interoperability when defined across distinct blockchains. Technological limitations such as different capabilities in terms of available cryptographic mechanisms, including impossibility to verify signatures, (zk-)proofs, etc., also because of distinct capabilities at the blockchain core and wallets, represent a challenge \cite{yildiz2023toward}.

\smallskip
\textbf{General cross-chain communication challenges.}
% The previously mentioned challenges are specific to SSI. However, general cross-chain communication challenges also apply to the SSI case. In particular, cross-chain interoperation is mostly complex between heterogeneous blockchains \cite{zhong2021jointcloud,blaize_inter}. Scalability and speed remain open issues, particularly in light of the lack of standardization for cross-chain connectivity. Security concerns increase as any adopted cross-chain solution adds new attack vectors. Examples include concerns about data privacy and integrity - there is a risk of exposing sensitive information and tampering with the data, respectively, during the process. The user experience could be disappointing, e.g., if the user is forced to use more than one wallet \cite{blaize_inter}. Finally, while cross-chain solutions sometimes rely on centralized entities (e.g., bridge operations or oracles), it is essential to preserve decentralization.
In addition to SSI-specific issues, general cross-chain challenges also apply. Interoperability between heterogeneous blockchains is complex \cite{zhong2021jointcloud,blaize_inter}, with scalability, speed, and lack of standards remaining open problems. Security risks include data privacy and integrity, while user experience can suffer if multiple wallets are required \cite{blaize_inter}. Maintaining decentralization is also critical, as some cross-chain solutions rely on centralized entities like bridges or oracles.

%-------------------------------------------
%-------------------------------------------
%-------------------------------------------
\subsection{Specific Challenges for the Identified Scenarios}
\label{sec:challanges_specific}

Table \ref{tab:challanges} exemplifies specific challenges in relation to the scenarios identified in Section \ref{sec:scenarios}. 
%For easiness of exposure, the challenges are listed for the case $n=2$ (i.e., two blockchains) but can be trivially extended for any $n \geq 2$.

%---------------------------------
\begin{table*}
\begin{footnotesize}
%\begin{center}
\begin{tabular}{p{0.08\textwidth}p{0.16\textwidth}p{0.67\textwidth}}
\hline
\textbf{Scenario} & \textbf{Description} & \textbf{Specific Challenges} \\
\hline
A1-B1 & One single DID, on-chain verif. & C1.1 One blockchain reads information from another blockchain (e.g., $\mathsf{BC_2}$ reads DID related information from $\mathsf{BC_1}$) \\
& & C1.2 Both blockchains need to be reachable simultanously \\
%& \\
A2-B1 & One DID per blockchain, & C2.1 Automatic updates of the DID related information from one blockchain to the others (e.g., updates related to $\mathsf{DID_1}$ on $\mathsf{BC_2}, \dots, \mathsf{BC_n}$)\\
& on-chain verif. & C2.2  Unlinkability of the corresponding DIDs $\{\mathsf{DID_1}, \dots, \mathsf{DID_n}\}$\\
& & C2.3 No use of intermediaries, especially centralized (e.g., oracles) \\
& & C2.4 High interoperability between (heterogeneous) blockchains \\
& & C2.5 Duplication of identities\\
& & C2.6 Scalability for large $n$ (i.e., grow in the number of blockchains)\\
\hline 
B2 & Off-chain verif. & C3.1 Maintain decentralization \\
& & C3.2 Avoid to introduce intermediaries \\
& & C3.3 Avoid to introduce a new attack vector/point of trust \\
\hline
\end{tabular}
\centering
\end{footnotesize}
\caption{Specific challenges for the identified scenarios}
\label{tab:challanges}
\end{table*}
%---------------------------------

\smallskip
\textbf{Scenario A1-B1.} Reading from one blockchain directly into another is still not a trivial task. Moreover, the blockchain that reads the information, as well as the blockchain from which the information is read, must both be reachable simultaneously. 

\smallskip
\textbf{Scenario A2-B1.} 
% A public linking of the DIDs of the same owner in distinct blockchains diminishes privacy and might damage anonymity. Cross-chain identity should typically encompass knowledge of the state of DIDs on all blockchains (if they have been revoked, the DID methods have been updated, etc.). Revocation is already problematic in a single blockchain, so revocation in cross-chain SSI becomes even harder. Scalability for a large number of blockchains, i.e., a large $n$ might be problematic in this context. Similarly, the existing work showed to be challenging to build cross-chain identity without the use of intermediaries, \cite{zecchini2023building} claiming to be the first to accomplish this. Contrary to the first scenario, which only requires one-sided reading, the second needs higher inter-connectivity. Finally, while having one identity on each blockchain facilitates usability, it duplicates identities.  
Linking a user’s DIDs across multiple blockchains weakens privacy and anonymity. A cross-chain identity must track the state of all associated DIDs—revocations, rotations, and updates—making management complex. Revocation is already difficult on a single blockchain and even harder across many, raising scalability and synchronization issues. Building cross-chain SSI without intermediaries remains challenging, with \cite{zecchini2023building} among the first to address it for permissionless blockchains. While having separate DIDs per chain improves usability, it introduces redundancy and weakens the notion of a unified SSI.

\smallskip
\textbf{Scenarios A1-B2 and A2-B2.} We consider the same set of specific challenges for the scenarios with off-chain verification. The placement of the verifier off-chain diminishes decentralization. Although the verifier could be implemented in a decentralized manner, it will not achieve a similar layer as a decentralized ledger. Also, by construction, it becomes an intermediary, introducing a new vector of attacks and assuming a new point of trust.

%-------------------------------------------
%-------------------------------------------
%-------------------------------------------
\section{Analysis of Different Cross-Chain Mechanisms}
\label{sec:sol_types}

%-------------------------------------------
%-------------------------------------------
%-------------------------------------------
%-------------------------------------------
%---------------------------------
\begin{table*}
\begin{footnotesize}
\begin{center}
\begin{tabular}{p{0.15\textwidth}p{0.24\textwidth}p{0.23\textwidth}p{0.23\textwidth}}
\hline
\textbf{Model} & \textbf{Short description}  & \textbf{Strengths} & \textbf{Weaknesses/Limitations} \\
\hline
Light-Client Verification \cite{light_client_paper1,light_client_paper2,ghosh2021decentralized} & Each chain runs a lightweight client of the other chain, and consensus proofs are directly verified \cite{light_client_paper1} & Trustless and verifiable across chains; security based on consensus proofs; native, cryptographic interoperability, no need for intermediaries \cite{ghosh2021decentralized,light_client_paper1,light_client_paper2} & High gas cost and computational for proof verification, limited scalability \cite{ghosh2021decentralized,light_client_paper1,light_client_paper2}\\
%---------------------
Notary / Federated Attestation \cite{tran2024crosscert} & A trusted set of notaries observes one chain and attests identity events on the other chain \cite{tran2024crosscert} & Governance; low-cost, efficient, interoperability between public and permissioned ledgers \cite{tran2024crosscert} & Relies on notaries; weaker decentralization \cite{tran2024crosscert}\\
%---------------------
Hub-and-Spoke Architecture \cite{zecchini2023building} & A central identity hub mediates interoperability among multiple spoke chains, providing a unified resolver or trust registry \cite{zecchini2023building} & Central interoperability hub; enforces policy and governance consistency; simplified DID and VC resolution \cite{zecchini2023building} & Single point of failure (the hub); complex governance and scalability; not fully decentralized \cite{zecchini2023building}\\
%---------------------
Hash Time-Locked Contracts (HTLCs)-based SSI \cite{tran2024crosscert} & Uses cryptographic time-locked swaps to atomically exchange identity proofs or credentials between two chains without intermediaries \cite{tran2024crosscert} & Trustless and atomic; no intermediaries; privacy-preserving credential exchange \cite{tran2024crosscert} & Binary logic, limited flexibility; SSI complex setup; requires synchronous or atomic exchanges \cite{tran2024crosscert}\\
%---------------------
General Message Passing (Relayer/Guardian) \cite{general_message_passing} & A third-party relayer forwards signed identity events and proofs between independent chains \cite{general_message_passing} & Fast deployment; supports different chain architectures; low and adaptable on-chain footprint \cite{general_message_passing} & Trust in relayers/guardians, security depends on the honesty of parties; relayer set compromises security \cite{general_message_passing}\\
%---------------------
Off-Chain Verification + On-Chain Attestation \cite{off_chain_verif} & DID and credential operations occur off-chain, with only cryptographic hashes or proofs anchored on one or more blockchains \cite{off_chain_verif} & High scalability and privacy; low cost; enables multi-chain anchoring and zk-proofs \cite{off_chain_verif} & Requires trusted off-chain resolvers; complex proof synchronization; the verification transparency is dependent on the off-chain logic \cite{off_chain_verif}\\
%---------------------
Side-Chain Model \cite{tran2024crosscert} & Transactions are run on a specialized side-chain anchored to a main chain \cite{tran2024crosscert} & Scalable and cost-efficient operations; logic and governance supports SSI; periodic anchoring provides linkage \cite{tran2024crosscert} & Bridge security; possible weaker side-chain consensus; replay and state desynchronization risk \cite{tran2024crosscert}\\
%---------------------
Relay-Chain Model \cite{tran2024crosscert} & A shared relay chain coordinates multiple dependent para-chains, enabling native cross-chain identity communication \cite{tran2024crosscert} & Shared security; native interoperability; enables ecosystem-scale SSI systems \cite{tran2024crosscert} & Assumes chains have the same ecosystem; governance complexity; hard to connect external (non-relay) blockchains \cite{tran2024crosscert}\\
%---------------------
Multi-Anchor / Multi-Ledger DID & A DID is anchored on multiple blockchains, ensuring redundancy and cross-ledger resolvability; e.g., a \textit{multi-chain DID Registry}, with a single registry replicated across multiple chains %, so that each chain can resolve DIDs without querying another chain 
\cite{li2025blockchain} & Resilient and redundant; portable across ecosystems; increase interoperability & Complex synchronization; higher cost for multiple anchors; inconsistency risk\\
%---------------------
Cross-Chain Credential Portability & Credentials remain off-chain but can be verified on any chain using standard signatures and DID methods; e.g., \textit{off-chain resolver service}, with third-parties or decentralized nodes to maintain DID resolution for multiple chains \cite{li2025blockchain} & Chain-agnostic verification & Difficult revocation and update sync; consistent verification context across ecosystems; depends on DID resolution compatibility\\
\hline
\end{tabular}
\end{center}
\end{footnotesize}
\caption{Cross-chain for SSI}
\label{tab:cross-chain_idm}
\end{table*}
%---------------------------------
%---------------------------------
%---------------------------------
\subsection{Cross-Chain for IdM}
\label{sec:cc_idm}

We identify several cross-chain models for IdM and investigate their strengths and weaknesses. Table \ref{tab:cross-chain_idm} lists our findings. The models can fit one or more scenarios, as defined in Section \ref{sec:scenarios}. For example, the Light-Client Verification is a candidate for Scenario~A1-B1 (single DID, on-chain verification); the Off-Chain Verification + On-Chain Attestation fits any of the two B2 scenarios (off-chain verification); Hub-and-Spoke Architectures fit best with Scenario A2-B1 (multiple DIDs, on-chain verification).
%We identified the following cross-chain IdM models: \textit{Light-Client Verification}, \textit{Notary / Federated Attestation}, \textit{Hub-and-Spoke Architecture}, \textit{Hash Time-Locked Contracts}, \textit{General Message Passing (Relayer/Guardian)}, \textit{Off-Chain Verification + On-Chain Attestation}, \textit{Side-chain Model}, \textit{Relay-Chain Model}, \textit{Multi-Anchor / Multi-Ledger DID}, \textit{Cross-Chain Credential Portability}.

%-------------------------------------------
\subsection{Projects and Industry Solutions}
\label{sec:cc_sol}

We further explore projects and industry solutions that develop VCs and decentralized identity frameworks, offering practical tools and implementations. 

\textit{Hypersign} \cite{hypersign} offers a privacy-preserving, enterprise-compliant identity and access management platform with selective disclosure. 
\textit{ACA-Py}, \textit{CREDEBL}, and \textit{Hyperledger Identus}, hosted under the LF Decentralized Trust initiative, deliver agent-based frameworks and infrastructure for SSI, with a focus on interoperability through the \textit{Aries} protocols~\cite{identus}. Consortia and governments widely use these tools for pilots of digital identity wallets, and they provide libraries and SDKs that simplify the development of interoperable identity solutions.
\textit{BitXHub} \cite{bitxhub} from Hyperchain enables cross-chain identity synchronization via relay chains, offers a modular architecture for cross-chain communication, and has been adopted in financial and government scenarios \cite{chinapractical}.
The \textit{Inter-Blockchain Communication Protocol (IBC)} and community projects such as the \textit{DoraHacks}~\cite{dorahacks} seek to extend verifiable identity to multi-chain ecosystems. IBC is integrated within the \textit{Cosmos} ecosystem and is used to validate user credentials across heterogeneous chains. 
\textit{XuperChain} \cite{xuperchain} supports large-scale DID integration, while \textit{Tencent Cloud’s TDID} \cite{tencent} brings decentralized identity to enterprise cloud environments. 
The \textit{LF Decentralized Trust} directory \cite{trustproject} shows the diversity of solutions, and the \textit{Universal Resolver} \cite{univ_resolver} provides DID resolution across networks.
The existing projects highlight the growing interest in decentralized trust infrastructures. 

%Andreea
%We review industry projects developing verifiable credentials and decentralized identity solutions. Hypersign \cite{hypersign} offers a privacy-preserving identity and access platform with selective disclosure for enterprise compliance. 
%\textit{ACA-Py}, \textit{CREDEBL}, and \textit{Hyperledger Identus} provide agent-based SSI frameworks with \textit{Aries} protocols for interoperability, widely used in digital identity wallet pilots \cite{identus}. 
%\textit{BitXHub} \cite{bitxhub} enables cross-chain identity synchronization via relay chains, adopted in finance and government \cite{chinapractical}. 
%The \textit{Inter-Blockchain Communication Protocol (IBC)} and projects like \textit{DoraHacks} \cite{dorahacks} extend verifiable identity across heterogeneous chains within the \textit{Cosmos} ecosystem. 
%\textit{XuperChain} \cite{xuperchain} supports large-scale DID integration, while \textit{Tencent Cloud’s TDID} \cite{tencent} brings decentralized identity to enterprise cloud environments. 
%The LF Decentralized Trust directory \cite{trustproject} highlights the diversity of SSI solutions, and the Universal Resolver \cite{univ_resolver} provides DID resolution across networks.

%-------------------------------------------
%-------------------------------------------
%-------------------------------------------
\section{Discussion}
\label{sec:disc}

%-------------------------------------------
%-------------------------------------------
%-------------------------------------------
\subsection{On the Utility of Cross-Chain SSI}

Despite strong interest in large-scale digital identity and blockchain’s decentralization benefits, blockchain-based SSI has seen limited adoption \cite{panait2020analysis}. This is partly because associating identity attributes with blockchain addresses conflicts with the original concept of blockchain. Still, blockchain SSI offers advantages: simpler key management than PKI, easy signature verification, and, with cross-chain SSI, improved interoperability, portability, and resilience against downtime, forks, or governance issues. However, the interconnectivity of multiple blockchains raises governance and interoperability questions: \textit{Which networks should be trusted to anchor public keys? How are conflicts resolved if issuers update or revoke their keys on one chain but not another? How do verifiers determine the data sources? How can cross-chain SSI be implemented efficiently and scalable?} In the end, blockchains in general and cross-chain SSI in particular introduce a complexity that must be well motivated and assumed.

%-------------------------------------------
%-------------------------------------------
%-------------------------------------------
\subsection{Security and Privacy}

% Many security issues arise from the inappropriate understanding and usage of concepts and technology. The blockchain is currently a candidate for anchoring DIDs, but it must not be perceived as a storage solution for VCs. However, there are still solutions (e.g., \cite{zhong2021jointcloud}) that propose VC storage on the chain. VCs contain sensitive information that should never be stored on-chain, as already discussed in Section \ref{sec:challanges_general}. On the other side, due to the immutability property of blockchain, a VC stored on-chain cannot be altered, and thus the integrity of the information is preserved \cite{misc_protokol}. To accommodate this without full compromise of security and privacy, the hash of a VC could be stored on-chain (refer to Section \ref{sec:back_ssi}).

Security issues often stem from the misuse of technology. While blockchains are suitable for anchoring DIDs, they should not store VCs containing sensitive data, though some solutions still do \cite{zhong2021jointcloud}. On-chain storage preserves integrity due to immutability \cite{misc_protokol}, but a safer approach is storing only the hash of a VC on-chain (Section \ref{sec:back_ssi}).

GDPR compliance requires holders to delete or revoke DIDs and VCs, which conflicts with blockchain’s immutable storage \cite{gdpr,w3c_did,w3c_did_draft,ebsi_white_paper}. Pairwise DIDs—known only to the holder and one other party—help preserve privacy, with n-wise DIDs extend this to multiple parties \cite{yildiz2023toward,dif_pairwise_did}.
In cross-chain SSI, revocation is even more challenging: updates must propagate across chains, risking desynchronization and outdated data, while the DID history remains on-chain, raising privacy concerns \cite{yildiz2023toward}. Current works rarely address cross-chain revocation in detail.
%--------------------------------------------------------
%So, the following open question arises: \textit{How can information update and, in particular, revocation can be implemented efficiently and scalable in cross-chain SSI?}  \ruxandra{TODO - scenario A2}

% Blockchains strongly rely on asymmetric cryptography. It is known that many asymmetric solutions are susceptible to quantum attacks. Most of the existing and recently introduced blockchain-based SSI solutions fail to implement post-quantum resistance, and even keep it out of their goals. Only a few works look into the transition to post-quantum SSI \cite{solavagione2025transition,qubip,zeydan2024post}. In the case of cross-chain, if one chain is susceptible to quantum attacks, it lowers the overall security and introduces new attack points on quantum-resistant blockchain solutions (the principle of the weakest link).

Blockchains rely on asymmetric cryptography, which is vulnerable to quantum attacks. Most blockchain-based SSI solutions lack post-quantum resistance, though a few address it \cite{solavagione2025transition,qubip,zeydan2024post}. In cross-chain settings, a quantum-vulnerable chain weakens overall security, creating new attack points.

% \textit{Unlinkability}, as defined in \cite{zecchini2023building}, ensures that two identifiers $DID_1$ and $DID_2$ defined on two distinct blockchains $BC_1$ and $BC_2$, respectively, cannot be linked to belong to the same holder unless the holder declares control of them both. 
% Nevertheless, once the holder needs to use the facilities given by the cross-chain SSI, then they become (publicly) linked. 
% %\ruxandra{TODO} Unlinkability only holds if the user does not link the two identities, so if the facility is not used. Once the user needs to somehow use them together they become publicly linked, i.e., everyone knows they belong to the same user. 
% If verification or other actions (e.g., updates, cross-data transfer) are performed on-chain using a smart contract and the two identifiers are exposed, then unlinkability gets broken.

% One limitation of cross-chain SSI solutions that follow Scenario A2, such as \cite{zecchini2023building}, is automatic information propagation, in the sense that if one holder changes the identifier information in one blockchain, the change does not automatically propagate to all other blockchains. One solution would be to handle this by triggering a transaction to one smart contract that operates on each blockchain.  However, the management of these smart contracts can be expensive.
Unlinkability ensures that two DIDs on separate blockchains cannot be linked to the same holder unless the holder chooses to reveal it \cite{zecchini2023building}. However, using cross-chain SSI services can publicly link them, especially if on-chain actions expose both identifiers.

%A limitation of Scenario A2 cross-chain SSI is that updates to one DID do not automatically propagate to others. Propagating changes via smart contracts is possible, but can be costly to manage.

%-------------------------------------------
%-------------------------------------------
%-------------------------------------------
\subsection{Other Aspects}

DIDs can be modeled with a single key per document, as in EBSI’s did:key method \cite{ebsi_did}, which is ledger-independent and interoperable \cite{w3c_did_methods}. Multiple keys per DID are allowed, enabling cross-chain SSI solutions to include several identifiers in one document \cite{zecchini2023building}, though this exposes links and may violate privacy.

Cross-chain SSI faces key challenges in efficiency, portability, and scalability. Efficiency can be improved using layer-2 solutions, which handle DID operations off the main blockchain, reducing costs and increasing speed \cite{sidetree}. Portability requires standardization and interoperable tools, such as the universal wallet, allowing users to manage and use their identities seamlessly across different blockchains \cite{yildiz2023toward}. Scalability remains a limitation due to blockchain constraints, including throughput and congestion \cite{algorand}. Cross-chain SSI can mitigate these issues by enabling identity transfer between blockchains, maintaining continuity, and supporting resilient multi-chain IdM while preserving security and privacy.
%-------------------------------------------
%-------------------------------------------
%-------------------------------------------
\section{Conclusions}
\label{sec:conclusion}

The paper addresses the blockchain isolation problem in blockchain-based SSI. Although SSI presents an interest nowadays, the context of cross-chain SSI suffers from a lack of investigation. We classify cross-chain methods based on two criteria (number of DIDs and the placement of the verifier on-chain vs. off-chain), define specific requirements, and identify challenges. We evaluate several cross-chain methods and solutions (both theoretical models and practical solutions) and relate them to the context of cross-chain SSI. Nevertheless, our work identifies open problems and outlines future research directions.

\section*{\uppercase{Acknowledgements}}
This work was supported by a grant of the Ministry of Research, Innovation and Digitalization, CNCS/CCCDI - UEFISCDI, project number ERANET-CHISTERA-IV-PATTERN, within PNCDI IV.
% If any, should be placed before the references section without numbering. To do so please use the following command: 
% \textit{$\backslash$section*\{ACKNOWLEDGEMENTS\}}

\bibliographystyle{apalike}
{\small
\bibliography{icissp.bib}}

@inproceedings{tran2024crosscert,
  title={CrossCert: A privacy-preserving cross-chain system for educational credential verification using zero-knowledge proof},
  author={Tran, Tuan-Dung and Minh, Phong Khuu and Thuy, Trang Le Tran and Duy, Phan The and Cam, Nguyen Tan and Pham, Van-Hau},
  booktitle={International Conference on Industrial Networks and Intelligent Systems},
  pages={256--271},
  year={2024},
  organization={Springer}
}

@inproceedings{liu2022research,
  title={Research on cross-chain method based on distributed Digital Identity},
  author={Liu, Sihan and Mu, Tong and Xu, Shicheng and He, Guangyu},
  booktitle={Proceedings of the 2022 4th International Conference on Blockchain Technology},
  pages={59--73},
  year={2022}
}

@inproceedings{zecchini2023building,
  title={Building a cross-chain identity: A self-sovereign identity-based framework},
  author={Zecchini, Marco and Sober, Michael and Schulte, Stefan and Vitaletti, Andrea},
  booktitle={2023 IEEE International Conference on Decentralized Applications and Infrastructures (DAPPS)},
  pages={149--156},
  year={2023},
  organization={IEEE}
}

@inproceedings{zhong2021jointcloud,
  title={JointCloud cross-chain verification model of decentralized identifiers},
  author={Zhong, Tao and Shi, Peichang and Chang, Junsheng},
  booktitle={2021 IEEE International Performance, Computing, and Communications Conference (IPCCC)},
  pages={1--8},
  year={2021},
  organization={IEEE}
}

@inproceedings{gruner2021analyzing,
  title={Analyzing interoperability and portability concepts for self-sovereign identity},
  author={Gr{\"u}ner, Andreas and M{\"u}hle, Alexander and Meinel, Christoph},
  booktitle={2021 IEEE 20th International Conference on Trust, Security and Privacy in Computing and Communications (TrustCom)},
  pages={587--597},
  year={2021},
  organization={IEEE}
}

@article{yildiz2023toward,
  title={Toward interoperable self-sovereign identities},
  author={Yildiz, Hakan and K{\"u}pper, Axel and Thatmann, Dirk and G{\"o}nd{\"o}r, Sebastian and Herbke, Patrick},
  journal={IEEE Access},
  volume={11},
  pages={114080--114116},
  year={2023},
  publisher={IEEE}
}

@incollection{broshka2024evaluating,
  title={Evaluating the Importance of SSI-Blockchain Digital Identity Framework for Cross-Border Healthcare Patient Record Management},
  author={Broshka, Esmeralda and Jahankhani, Hamid},
  booktitle={Navigating the Intersection of Artificial Intelligence, Security, and Ethical Governance: Sentinels of Cyberspace},
  pages={87--110},
  year={2024},
  publisher={Springer}
}

@inproceedings{naghmouchi2021automatized,
  title={An automatized identity and access management system for IoT combining self-sovereign identity and smart contracts},
  author={Naghmouchi, Montassar and Ayed, Hella Kaffel Ben and Laurent, Maryline},
  booktitle={International symposium on foundations and practice of security},
  pages={208--217},
  year={2021},
  organization={Springer}
}

@misc{misc_protokol,
    author = {{Protokol}},
    title = {{Enhancing Trust with Blockchain Verifiable Credentials}},
    howpublished = {\url{https://www.protokol.com/insights/blockchain-verifiable-credentials}},
    year = {2025},
    note = {Last accessed: August, 2025}
}

@inproceedings{ghosh2021decentralized,
  title={Decentralized cross-network identity management for blockchain interoperation},
  author={Ghosh, Bishakh Chandra and Ramakrishna, Venkatraman and Govindarajan, Chander and Behl, Dushyant and Karunamoorthy, Dileban and Abebe, Ermyas and Chakraborty, Sandip},
  booktitle={2021 IEEE International Conference on Blockchain and Cryptocurrency (ICBC)},
  pages={1--9},
  year={2021},
  organization={IEEE}
}

@inproceedings{xie2022cross,
  title={Cross-Chain-Based Distributed Digital Identity: A Survey},
  author={Xie, Tianxiu and Zhang, Hong and Feng, Yiwei and Qi, Jing and Guo, Chennan and Yang, Gangqiang and Gai, Keke},
  booktitle={International Conference on Smart Computing and Communication},
  pages={525--534},
  year={2022},
  organization={Springer}
}

@article{panait2020identitymanagementblockchain,
  title={{Identity Management on Blockchain--Privacy and Security Aspects}},
  author={Panait, Andreea-Elena and Olimid, Ruxandra F and Stefanescu, Alin},
  journal={Proceedings of the Romanian Academy, Series A},
  volume = {21},
  number = {1},
  year = {2020},
  pages={45--52}
}

@article{dragnoiu2024identitymanagementsolutionarweave,
      title={{Towards an identity management solution on Arweave}}, 
      author={Andreea Elena Drăgnoiu and Ruxandra F. Olimid},
      year={2024},
      eprint={2412.13865},
      archivePrefix={arXiv},
      primaryClass={cs.CR},
      url={https://arxiv.org/abs/2412.13865}, 
}

@inproceedings{panait2020analysis,
  title={{Analysis of uPort Open, an identity management blockchain-based solution}},
  author={Panait, Andreea-Elena and Olimid, Ruxandra F and Stefanescu, Alin},
  booktitle={International Conference on Trust and Privacy in Digital Business},
  pages={3--13},
  year={2020},
  organization={Springer}
}

@article{boulet2025digital,
  title={Digital transformation of certificates issued by universities for European competitiveness},
  author={Boulet, Pierre and de Co{\"e}tlogon, Perrine and Th{\'e}bault, {\'E}milie and van de Velde, Patrice and Murie, Cyril},
  year={2025}
}

@misc{ebsi,
    author = {{European Commission}},
    title = {{ European Blockchain Services Infrastructure (EBSI)}},
    howpublished = {\url{https://ec.europa.eu/digital-building-blocks/sites/spaces/EBSI/pages/447687044/Home}},
    note = {Last accessed: August, 2025},
    year = {2025}
}

@misc{ebsi_vc,
    author = {{European Commission - EBSI}},
    title = {{Verifiable Credentials Framework}},
    howpublished = {\url{https://ec.europa.eu/digital-building-blocks/sites/display/EBSI/EBSI+Verifiable+
                         Credentials}},
    note = {Last accessed: August, 2025},
    year = {2025}
}

@article{tan2023verification,
  title={Verification of education credentials on European Blockchain Services Infrastructure (EBSI): action research in a cross-border use case between Belgium and Italy},
  author={Tan, Evrim and Lerouge, Ellen and Du Caju, Jan and Du Seuil, Dani{\"e}l},
  journal={Big Data and Cognitive Computing},
  volume={7},
  number={2},
  pages={79},
  year={2023},
  publisher={MDPI}
}

@misc{medium_astrakode,
    author = {{Medium - AstraKode}},
    title = {{Digital Identity and Blockchain: Exploring the EU Digital Identity Framework}},
    howpublished = {\url{https://astrakode.medium.com/digital-identity-and-blockchain-exploring-the-eu-digital-identity-framework\-cdec759e03ba}},
    note = {Last accessed: August, 2025},
    year = {2025}
}

@misc{blaize_inter,
    author = {{Blaize - S. Onyshchenko}},
    title = {{Cross-Chain Interoperability: Unlocking the Power of Blockchain}},
    howpublished = {\url{https://blaize.tech/blog/cross-chain-interoperability}},
    note = {Last accessed: August, 2025},
    year = {2024}
}

@misc{w3c_did,
  author = {{W3C Recommendation}},
  title = {{Decentralized Identifiers (DIDs) v1.0. Core architecture, data model, and representations}},
  year = 2022,
  note = {\url{https://www.w3.org/TR/did-core}. Last Accessed: August 2025}
}

@misc{w3c_did_methods,
  author = {{W3C Group Note}},
  title = {{DID Methods - Known DID Methods in the Decentralized Identifier Ecosystem}},
  year = 2025,
  note = {\url{https://www.w3.org/TR/did-extensions-methods}. Last Accessed: September 2025}
}

@misc{w3c_did_draft,
  author = {{W3C Recommendation}},
  title = {{Decentralized Identifiers (DIDs) v1.1. Core architecture, data model, and representations}},
  year = 2025,
  note = {\url{https://www.w3.org/TR/did-1.1}. Last Accessed: August 2025}
}

@misc{w3c_verifcred,
  author = {{W3C Recommendation}},
  title = {{Verifiable Credentials Data Model v2}},
  note = {\url{https://www.w3.org/TR/vc-data-model-2.0}. Last Accessed: August 2025},
  year = {2025}  
}

@misc{eu_age_solution,
  author = {{European Commission}},
  title = {{European Age Verification Solution}},
  note = {\url{https://ageverification.dev}. Last Accessed: August 2025},
  year = 2025
}

@misc{eu_id_wallet,
  author = {{European Commission}},
  title = {{EU Digital Identity Wallet}},
  note = {\url{https://ec.europa.eu/digital-building-blocks/sites/spaces/EUDIGITALIDENTITYWALLET
               /pages/694487738/EU+Digital+Identity+Wallet
               +Home}. Last Accessed: August 2025},
  year = 2025
}

@misc{dif,
  author = {{Decentralized Identity Foundation}},
  title = {{DIF}},
  note = {\url{https://identity.foundation}. Last Accessed: August 2025},
  year = 2025
}

@misc{dif_faq,
  author = {{Decentralized Identity Foundation}},
  title = {{Decentralized Identity FAQ}},
  note = {\url{https://identity.foundation/faq}. Last Accessed: August 2025},
  year = 2025
}

@inproceedings{sidhu2022trust,
  title={Trust development for blockchain interoperability using self-sovereign identity integration},
  author={Sidhu, Sahilpreet Singh and Nguyen, Minh Nam Hai and Ngene, Chikamnaele and Rouhani, Sara},
  booktitle={2022 IEEE 13th Annual Information Technology, Electronics and Mobile Communication Conference (IEMCON)},
  pages={0033--0040},
  year={2022},
  organization={IEEE}
}

@article{li2025blockchain,
author = {Li, Ningran and Qi, Minfeng and Xu, Zhiyu and Zhu, Xiaogang and Zhou, Wei and Wen, Sheng and Xiang, Yang},
title = {Blockchain Cross-Chain Bridge Security: Challenges, Solutions, and Future Outlook},
year = {2025},
issue_date = {March 2025},
publisher = {Association for Computing Machinery},
address = {New York, NY, USA},
volume = {4},
number = {1},
url = {https://doi.org/10.1145/3696429},
doi = {10.1145/3696429},
journal = {Distrib. Ledger Technol.},
month = feb,
articleno = {8},
numpages = {34}
}

@misc{ewc_travel,
    author = {{EU Digital Wallet Consortium (EWC)}},
    title = {{LinkedIn post}},
    howpublished = {\url{https://www.linkedin.com/posts/eu-digital-identity-wallet-consortium-ewc_eudiw-digitalidentity-travelinnovation-activity-7360661789686042624-HjYw}},
    note = {Last accessed: September, 2025},
    year = 2025
}

@misc{ec_use_cases,
    author = {{European Commission}},
    title = {{Pilot projects - The many use cases of EU Digital Identity Wallets}},
    howpublished = {\url{https://ec.europa.eu/digital-building-blocks/sites/spaces/EUDIGITALIDENTITYWALLET
                    /pages/716146139/The+many+use+cases+of+the+EU
                         +Digital+Identity+Wallet}},
    note = {Last accessed: September, 2025},
    year = 2025
}

@article{ou2022overview,
  title={An overview on cross-chain: Mechanism, platforms, challenges and advances},
  author={Ou, Wei and Huang, Shiying and Zheng, Jingjing and Zhang, Qionglu and Zeng, Guang and Han, Wenbao},
  journal={Computer Networks},
  volume={218},
  pages={109378},
  year={2022},
  publisher={Elsevier}
}

@inproceedings{augusto2024sok,
  title={Sok: Security and privacy of blockchain interoperability},
  author={Augusto, Andr{\'e} and Belchior, Rafael and Correia, Miguel and Vasconcelos, Andr{\'e} and Zhang, Luyao and Hardjono, Thomas},
  booktitle={2024 IEEE Symposium on Security and Privacy (SP)},
  pages={3840--3865},
  year={2024},
  organization={IEEE}
}

@inproceedings{zhang2024security,
  title={Security of cross-chain bridges: Attack surfaces, defenses, and open problems},
  author={Zhang, Mengya and Zhang, Xiaokuan and Zhang, Yinqian and Lin, Zhiqiang},
  booktitle={Proceedings of the 27th International Symposium on Research in Attacks, Intrusions and Defenses},
  pages={298--316},
  year={2024}
}

@article{belchior2024brief,
  title={A brief history of blockchain interoperability},
  author={Belchior, Rafael and S{\"u}{\ss}enguth, Jan and Feng, Qi and Hardjono, Thomas and Vasconcelos, Andr{\'e} and Correia, Miguel},
  journal={Communications of the ACM},
  volume={67},
  number={10},
  pages={62--69},
  year={2024},
  publisher={ACM New York, NY, USA}
}

@misc{eidas_doc,
  author = {{European Union}},
  title = {{Regulations of the European Parliament and of the Council amending Regulation (EU) No.910/2014 as regards establishing the European Digital Identity Framework}},
  year = 2023,
  note = {\url{https://data.consilium.europa.eu/doc/document/PE-68-2023-REV-1/en/pdf}. Last Accessed: August 2025}
}

@misc{gdpr,
  author = {{Regulation (EU) 2016/679}},
  title = {{General Data Protection Regulation GDPR}},
  note = {\url{https://gdpr-info.eu}. Last Accessed: August 2025},
  year = 2018
}

@misc{hyperledgerindy,
  author = {{LF Decentralized trust}},
  title = {{HyperledgerIndy}},
  note = {\url{https://www.lfdecentralizedtrust.org/projects
               /hyperledger-indy}. Last Accessed: September 2025},
  year = 2025
}

@misc{hyperledgerursa,
  author = {{LF Decentralized trust}},
  title = {{Hyperledger Ursa (EOL)}},
  note = {\url{https://lf-hyperledger.atlassian.net/wiki/spaces/ursa/overview}. Last Accessed: September 2025},
  year = 2025
}

@misc{hyperledgeraries,
  author = {{LF Decentralized trust}},
  title = {{Aries}},
  note = {\url{https://www.lfdecentralizedtrust.org/projects/aries}. Last Accessed: September 2025},
  year = 2025
}

@misc{ethereum,
  author = {{Ethereum}},
  note = {\url{https://ethereum.org}. Last Accessed: September 2025},
  year = 2025
}

@misc{ebsi_white_paper,
  author = {{EBSI}},
  title = {{Revocation by EBSI}},
  note = {\url{https://ec.europa.eu/digital-building-blocks/sites/spaces/EBSI/pages/693209363/
            What+to+do+when+good+Verifiable+Credentials
               +go+bad}. Last Accessed: September 2025},
  year = 2023
}

@misc{dif_pairwise_did,
  author = {{Decentralized Identity Foundation}},
  title = {{Peer DID Method Specification}},
  note = {\url{https://identity.foundation/peer-did-method-spec}. Last Accessed: September 2025},
  year = 2025
}

@inproceedings{solavagione2025transition,
  title={Transition of Self-Sovereign Identity to Post-Quantum Cryptography},
  author={Solavagione, Alberto and Vesco, Andrea},
  booktitle={2025 International Conference on Quantum Communications, Networking, and Computing (QCNC)},
  pages={174--181},
  year={2025},
  organization={IEEE}
}

@misc{qubip,
  author = {{{QUBIP}}},
  title = {{Post-Quantum Verifiable Credentials}},
  note = {\url{https://qubip.eu/post-quantum-verifiable-credentials}. Last Accessed: September 2025},
  year = 2024
}

@article{zeydan2024post,
  title={Post-quantum blockchain-based decentralized identity management for resource sharing in open radio access networks},
  author={Zeydan, Engin and Blanco, Luis and Mangues-Bafalluy, Josep and Arslan, Suayb S and Turk, Yekta},
  journal={IEEE Transactions on Green Communications and Networking},
  year={2024},
  publisher={IEEE}
}

@article{mazzocca2025survey,
  title={A survey on decentralized identifiers and verifiable credentials},
  author={Mazzocca, Carlo and Acar, Abbas and Uluagac, Selcuk and Montanari, Rebecca and Bellavista, Paolo and Conti, Mauro},
  journal={IEEE Communications Surveys \& Tutorials},
  year={2025},
  publisher={IEEE}
}

@article{mezquita2023blockchain,
  title={Blockchain-based supply chain systems, interoperability model in a pharmaceutical case study},
  author={Mezquita, Yeray and Podgorelec, Bla{\v{z}} and Gil-Gonz{\'a}lez, Ana Bel{\'e}n and Corchado, Juan Manuel},
  journal={Sensors},
  volume={23},
  number={4},
  pages={1962},
  year={2023},
  publisher={MDPI}
}

@misc{sidetree,
  author = {{Decentralized Identity Foundation}},
  title = {{Sidetree v1.0.1}},
  note = {\url{https://identity.foundation/sidetree/spec}. Last Accessed: September 2025},
  year = 2025
}

@misc{ebsi_did,
  author = {{EBSI}},
  title = {{DID Methods}},
  note = {\url{https://hub.ebsi.eu/vc-framework/did}. Last Accessed: September 2025},
  year = 2024
}

@misc{algorand,
  author = {{Algorand Technologies}},
  title = {{Solving the “Blockchain Trilemma”}},
  note = {\url{https://algorandtechnologies.com/technology/solving-the-blockchain-trilemma}. Last Accessed: September 2025},
  year = 2025
}

@misc{hypersign,
  title = {Hypersign},
  note = {\url{https://www.hypersign.id}. Last Accessed: September 2025},
  year = {2025}
}

@misc{bitxhub,
  title = {BitXHub},
  note = {\url{https://www.hyperchain.cn/en/products
               /bitxhub}. Last Accessed: September 2025},
  year = {2025}
}

@misc{chinapractical,
  title = {{Hyperchain development in China}},
  note = {\url{https://blog.csdn.net/Hyperchain/article/details/
               115204815}. Last Accessed: September 2025},
  year = {2025}
}

@misc{dorahacks,
  title = {{Creating a Crosschain DID on IBC}},
  note = {\url{https://dorahacks.io/buidl/18092}. Last Accessed: September 2025},
  year = {2025}
}

@misc{xuperchain,
  title = {XuperChain},
  note = {\url{https://xuper.baidu.com}. Last Accessed: September 2025},
  year = {2025}
}

@misc{tencent,
  title = {TDID},
  note = {\url{https://cloud.tencent.com/product/tdid}. Last Accessed: September 2025},
  year = {2025}
}

@misc{trustproject,
  title = {{LF Decentralized Trust Projects}},
  note = {\url{https://www.lfdecentralizedtrust.org/projects}. Last Accessed: September 2025},
  year = {2025}
}

@misc{identus,
  title = {Aries project},
  note = {\url{https://www.lfdecentralizedtrust.org/projects
               /aries}. Last Accessed: September 2025},
  year = {2025}
}

@misc{univ_resolver,
  author = {{Decentralized Identity Foundation}},
  title = {Universal Resolver},
  note = {\url{https://dev.uniresolver.io}. Last Accessed: September 2025},
  year = {2025}
}

@misc{cosmos,
  author = {{Cosmos}},
  title = {{The Inter-Blockchain Communication protocol}},
  note = {\url{https://cosmos.network/ibc}. Last Accessed: September 2025},
  year = {2025}
}

@ARTICLE{light_client_paper1,
  author={Paavolainen, Santeri and Carr, Christopher},
  journal={IEEE Access}, 
  title={Security Properties of Light Clients on the Ethereum Blockchain}, 
  year={2020},
  volume={8},
  number={},
  pages={124339-124358},
  doi={10.1109/ACCESS.2020.3006113}
}

@article{light_client_paper2,
title = {Efficient Query Verification for Blockchain Superlight Clients Using SNARKs},
journal = {Blockchain: Research and Applications},
pages = {100396},
year = {2025},
issn = {2096-7209},
doi = {https://doi.org/10.1016/j.bcra.2025.100396},
url = {https://www.sciencedirect.com/science/article/pii/S209672092500123X},
author = {Stefano {De Angelis} and Ivan Visconti and Andrea Vitaletti and Marco Zecchini}
}

@misc{general_message_passing,
  author = {{Messari}},
  title = {{Which Interoperability Protocol Leads the Way?}},
  year = 2025,
  note = {\url{https://messari.io/compare/layerzero-vs-axelar}. Last Accessed: September 2025}
}

@inbook{off_chain_verif,
author = {Henry, Tiphaine and Tucci-Piergiovanni, Sara},
year = {2024},
month = {09},
pages = {71-88},
title = {Secure Proof Verification Blockchain Patterns},
isbn = {978-3-031-70444-4},
doi = {10.1007/978-3-031-70445-1_5}
}
% \bibliographystyle{apalike}
% \bibliography{\jobname}

% \section*{\uppercase{Appendix}}

% If any, the appendix should appear directly after the references without numbering, and not on a new page. To do so please use the following command: \textit{$\backslash$section*\{APPENDIX\}}

\end{document}